\documentclass[twocolumn]{aastex631}

\usepackage{xcolor}  

\newcommand{\ttt}{\texttt}
\newcommand{\tess}{\textit{TESS}}
\usepackage[utf8]{inputenc}

\begin{document}

\title{TESSELLATE: Piecing Together the Variable Sky With \tess}

\author[0009-0001-6992-0898]{Hugh Roxburgh}\thanks{Co-lead authors; these authors contributed equally to this work.}
\affiliation{School of Physical and Chemical Sciences --- Te Kura Mat\={u}, University of Canterbury, Private Bag 4800, Christchurch 8140, New Zealand}
\affiliation{International Centre for Radio Astronomy Research, Curtin University, Bentley, WA 6102, Australia}

\author[0000-0003-1724-2885]{Ryan Ridden-Harper}\thanks{Co-lead authors; these authors contributed equally to this work.}
\affiliation{School of Physical and Chemical Sciences --- Te Kura Mat\={u}, University of Canterbury, Private Bag 4800, Christchurch 8140, New Zealand}
\correspondingauthor{Ryan Ridden-Harper} 
\email{ryan.ridden@canterbury.ac.nz}

\author[0009-0009-9143-1204]{Andrew Moore}
\affiliation{School of Physical and Chemical Sciences --- Te Kura Mat\={u}, University of Canterbury, Private Bag 4800, Christchurch 8140, New Zealand}

\author[0009-0008-4935-069X]{Clarinda Montilla}
\affiliation{School of Physical and Chemical Sciences --- Te Kura Mat\={u}, University of Canterbury, Private Bag 4800, Christchurch 8140, New Zealand}

\author[0009-0008-6765-5171]{Brayden Leicester}
\affiliation{School of Physical and Chemical Sciences --- Te Kura Mat\={u}, University of Canterbury, Private Bag 4800, Christchurch 8140, New Zealand}

\author[0009-0003-8380-4003]{Zachary G. Lane}
\affiliation{School of Physical and Chemical Sciences --- Te Kura Mat\={u}, University of Canterbury, Private Bag 4800, Christchurch 8140, New Zealand}

\author[0009-0006-7990-0547]{James Freeburn}
\affiliation{Centre for Astrophysics and Supercomputing, Swinburne University of Technology, John St, Hawthorn, VIC 3122, Australia}
\affiliation{ARC Centre of Excellence for Gravitational Wave Discovery (OzGrav), John St, Hawthorn, VIC 3122, Australia}
\author[0000-0002-4410-5387]{Armin Rest}
\affiliation{Space Telescope Science Institute, 3700 San Martin Drive, Baltimore, MD 21218, USA.}
\affiliation{The Johns Hopkins University, Baltimore, MD 21218, USA.}

\author[0000-0003-3257-4490]{Michele T. Bannister}
\affiliation{School of Physical and Chemical Sciences --- Te Kura Mat\={u}, University of Canterbury, Private Bag 4800, Christchurch 8140, New Zealand}

\author[0000-0002-5425-2655]{Andrew R. Ridden-Harper}
\affiliation{Independent researcher}

\author[0009-0005-0556-3886]{Lancia Hubley}
\affiliation{School of Physical and Chemical Sciences --- Te Kura Mat\={u}, University of Canterbury, Private Bag 4800, Christchurch 8140, New Zealand}
\author[0000-0001-5233-6989]{Qinan Wang}
\affiliation{Department of Physics and Kavli Institute for Astrophysics and Space Research, Massachusetts Institute of Technology, 77
Massachusetts Avenue, Cambridge, MA 02139, USA}
\author[0000-0002-0476-4206]{Rebekah Hounsell}
\affiliation{The University of Maryland Baltimore County, 1000 Hilltop Cir, Baltimore, MD 21250, USA}
\affiliation{NASA Goddard Space Flight Center, 8800 Greenbelt Road Greenbelt, MD 20771, USA}
\author[0000-0001-5703-2108]{Jeff Cooke}
\affiliation{Centre for Astrophysics and Supercomputing, Swinburne University of Technology, John St, Hawthorn, VIC 3122, Australia}
\affiliation{ARC Centre of Excellence for Gravitational Wave Discovery (OzGrav), John St, Hawthorn, VIC 3122, Australia}
\author[0000-0003-4263-2228]{Dave~A.~Coulter} 
\affiliation{Space Telescope Science Institute, 3700 San Martin Drive, Baltimore, MD 21218, USA.}
\author[0000-0002-9113-7162]{Michael M. Fausnaugh}
\affiliation{Department of Physics and Astronomy, Texas Tech University, Box 1051, Lubbock, TX 79409-1051, USA}

\begin{abstract}

We present \ttt{TESSELLATE}, a dedicated pipeline for performing an untargeted search documenting all variable phenomena captured by the \tess\ space telescope. 
Building on the \ttt{TESSreduce} difference imaging pipeline, \ttt{TESSELLATE} extracts calibrated and reduced photometric data for every full frame image in the \tess\ archive. 
Using this data, we systematically identify transient, variable and non-sidereal signals across timescales ranging from minutes to weeks. 
The high cadence and wide field of view of \tess\ enables us to conduct a comprehensive search of the entire sky to a depth of $\sim$17$\; m_i$. 
Based on the volumetric rates for known fast transients, we expect there to be numerous Fast Blue Optical Transients and Gamma Ray Burst afterglows present in the existing \tess\ dataset. 
Beyond transients, \ttt{TESSELLATE} can also identify new variable stars and exoplanet candidates, and recover known asteroids. 
We classify events using machine learning techniques and the work of citizen scientists via the Zooniverse \textit{Cosmic Cataclysms} project. 
Finally, we introduce the TESSELLATE Sky Survey: a complete, open catalog of the variable sky observed by \tess.

\end{abstract}


\section{Introduction}\label{sec:intro}

The \textit{Transiting Exoplanet Survey Satellite} \citep[\tess, ][]{Ricker2014} has imaged nearly the entire sky in a broad red to near-infrared band since starting operations in July 2018. Its wide field of view (FoV) of 96$\times$24 deg$^2$ and its continuous, high-cadence observation strategy have allowed \tess\ to fill a niche role in the transient community. 
While initially designed to hunt for exoplanets in the Milky Way, its science use has since expanded widely, including the study of variable stars \citep[][]{Bruch2022,Liu2023,Fetherolf2023}, supernovae \citep[SNe,][]{Fausnaugh2021,Vallely2021,Fausnaugh2023,Wang2023,Wang2024}, gamma-ray burst (GRB) afterglows \citep[][]{Smith2021,FausnaughGRB,Roxburgh2023,Jayaraman2024}, asteroids \citep{Pal2018,Pal2020,Woods2021, McNeill2023}, comets \citep{Ridden2021b,Farnham2021}, exocomets \citep{Zieba2019,Dobrycheva2023}, stellar flares \citep{Gunther20,Pietras2022}, and other transient phenomena. 

As the \tess\ mission has evolved, the volume of data produced from each target field, known as a \tess\ `sector', has greatly increased. This is due to the progressive shortening of its observation cadence: it began by capturing full-frame images (FFIs) every 30 minutes, shifted to a 10-minute cadence between 5\textsuperscript{th} July 2020 and 1\textsuperscript{st} September 2022, and has since been observing with a cadence of 200 seconds.

The data from each sector is released to the public through the MAST archive \citep{data:ffis} as a time-series of FFIs that are `calibrated' but not background subtracted. 
As such, observers are required to perform their own photometric extraction for sources that are not automatically extracted through the SPOC pipeline \citep[][]{Caldwell20}. 
There are a number of successful reduction pipelines available to the public 
e.g. TESSreduce \citep{Ridden2021}, the Quick Look Pipeline \citep[QLP;][]{Fausnaugh2021,Huang2020,Huang2020b,Fausnaugh2020qlp,Kunimoto2021,Kunimoto2022}, eleanor \citep{Feinstein2019}, and DIAmante \citep{DiIAmante}. Each pipeline has their own advantages and disadvantages due to unique choices of techniques.
However, these pipelines are often small-scale, focusing on producing cutouts of \tess\ data around desired coordinates or sources. 
This means that captured information has been left unexamined in the gaps between targeted searches.

With \tess's past success as a transient hunter across many time-domain fields, the motivation for a complete archival search of its images is clear. 
In particular, \tess\ is expected to be a powerful tool for observing fast extragalactic transients such as orphan GRB afterglows \citep{Freeburn2024,Perley2025}. 
Such events evolve on timescales of minutes to hours; the combination of \tess's wide field of view and rapid temporal resolution make it capable of discovering and sampling fast events at high time resolution.

In this paper, we introduce \texttt{TESSELLATE}: the \tess\ Extensive Lightcurve Logging and Analysis of Transient Events pipeline, the first system capable of automatically searching every \tess\ frame to document all fast transient events captured by \tess. 
For this pipeline, we use \ttt{TESSreduce} to remove the scattered light background and create difference images.
\texttt{TESSELLATE} is purpose-built for use on the OzSTAR supercomputing node held at Swinburne University, and is wrapped in a user-friendly \texttt{python} package\footnote{\url{DOI: https://doi.org/10.5281/zenodo.15954684}} that allows for simple but thorough control of the various processing steps. 

\begin{figure*}
    \centering
    \includegraphics[width=0.9\textwidth]{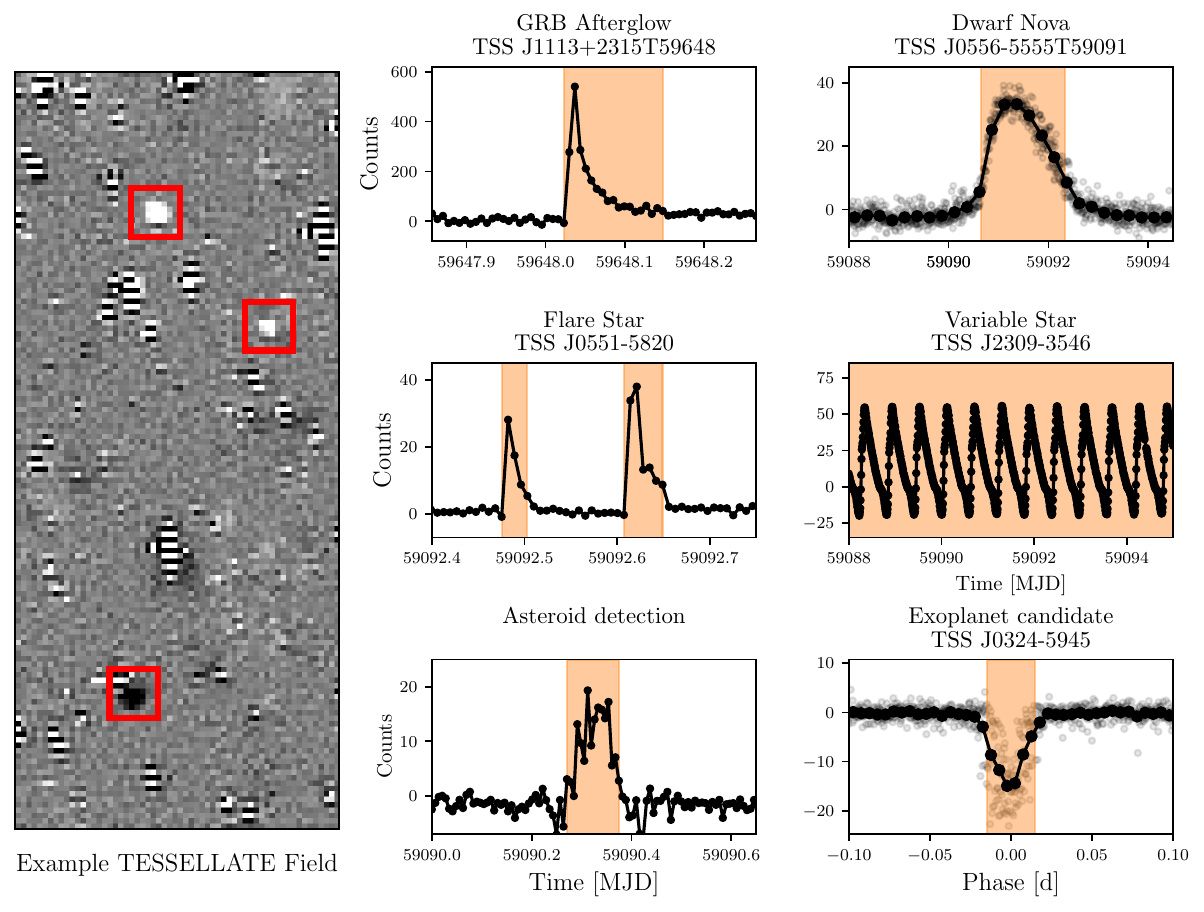}
     \caption{Event identification from \tess\ images with \ttt{TESSELLATE}. \textbf{Left:} Example image section that has been background subtracted and differenced. Sources identified with \ttt{TESSELLATE} are indicated with red boxes. \textbf{Center \& Right:} Example \tess\ light curves with events detected by \ttt{TESSELLATE} highlighted. For the recovered exoplanet (lower right) we show the phase-folded light curve with raw data (gray) and 10~minute binned light curve (black). Similarly for the dwarf nova (top right) we show the raw data (gray) and 6~hour binned light curve (black). The asteroid detection (lower left) shows an individual detection at a position on the detector and time.}
    \label{fig:main}
\end{figure*}

We developed \ttt{TESSELLATE} to be agnostic of transient type, allowing it to detect a wide range of variable, transient and non-sidereal phenomena. 
\autoref{fig:main} shows an example difference image with detections and an array of assorted event detections.
In this paper, all magnitudes are assumed to be AB unless stated otherwise, and magnitudes in \textit{TESS} are represented by $m_T$. 
Furthermore, we take $m_i\approx m_T$ for stellar spectral energy distributions, where $m_i$ is the apparent $i$ band magnitude.  
Section~\ref{sec:overview} describes each stage of the pipeline; Section~\ref{sec:eff-rates} provides detection efficiency estimates for a selection of \tess\ sectors; Section~\ref{sec:cases} highlights a variety of science cases to which \ttt{TESSELLATE} can be applied.
Finally, Section~\ref{sec:skysurvey} outlines the \ttt{TESSELLATE} Sky Survey made using TESSELATE, and how its data products are delivered.

\section{Pipeline Overview}
\label{sec:overview}

\begin{figure*}
    \centering
    \includegraphics[width=0.95\textwidth]{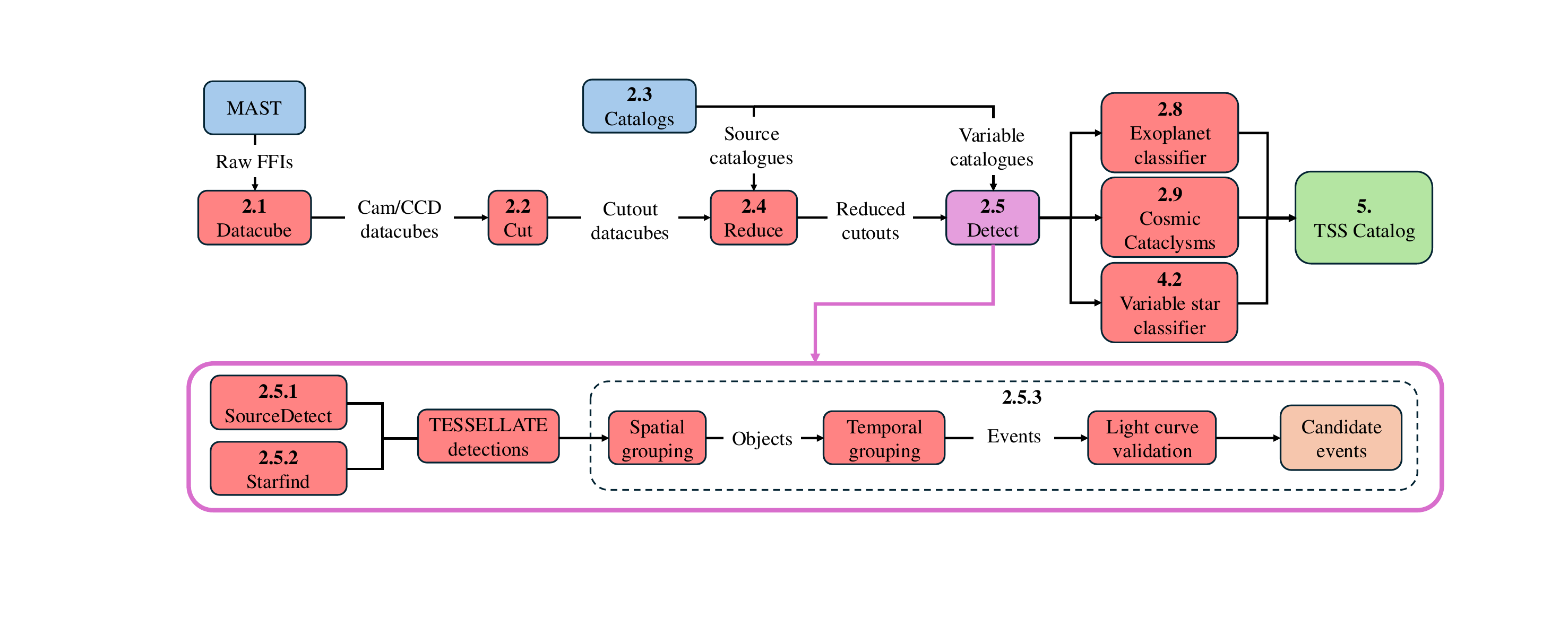}
    \caption{Schematic of the \ttt{TESSELLATE} pipeline. The detection workflow is expanded upon in the purple box. The section relevant to each step is indicated by the number at the top of the boxes.}
    \label{fig:workflow}
\end{figure*}

The \ttt{TESSELLATE} pipeline runs a full reduction and transient event extraction process on individual \tess\ sectors, simultaneously operating on all 16 CCDs from the 4 cameras of TESS. 
This pipeline is built upon the process described in \citet{Roxburgh2023} to conduct an untargeted transient search. 
When executed, TESSELLATE calls sub-processes sequentially, as visualized in \autoref{fig:workflow}.

We tested \ttt{TESSELLATE} using data from Sector~29, which observed from August 26$^{th}$ to September 21$^{st}$ 2020. 
This southern-pointing sector used the intermediate FFI cadence of 10~minutes and has significant complementary coverage with the DECam Local Volume Exploration Survey \citep[DELVE,][]{Drlica-Wagner2021}.

\subsection{TESS Datacube Generation}
\label{sec:data}
We acquire all TESS FFI data from the MAST archive. A full TESS image is constructed from four cameras with four CCDs. Each of the 16 CCDs form a 2048$\times$2048 pixel array with an angular resolution of 21" per pixel. Files are saved into hierarchical file structures based on the sector and camera information.

For each of the 16 CCDs, the FFIs are collated into three-dimensional datacubes termed `Target Pixel Files' (TPFs) using \texttt{astrocut} \citep{astrocut}. 
For sectors in the second extended mission, i.e. those with FFIs captured every 200 seconds, a single TPF datacube exceeds 300 GB. 
While they can be generated in this form, we divide these sectors into two datacubes. 
This split is made at the downlink time in \tess's orbit, which generally occurs near the halfway mark of the time series but can vary by a week either side.

\subsection{Cutting Datacubes}

In this stage, each of the generated datacubes are further divided into a number of cutouts, also completed using the \ttt{astrocut} package. This division enables further parallel processing and decreases the size of individual reduction tasks. Furthermore, the \ttt{TESSreduce} package produces more reliable subtractions of the scattered light background for smaller regions with less dynamic range in the background. The number of cutouts is adjustable; by default each datacube is divided into 16 equal cutouts of $512\times512\;$pixels.

\subsection{Catalog Acquisition} \label{sec:var_cat}
\ttt{TESSELLATE} gathers a number of external source catalogs to be used in data reduction and source identification. Most catalogs are accessed through Vizier queries made using \texttt{astroquery} \citep{astroquery}. For a given cutout, each query is centered on the central coordinates of the cutout, with a radius to fully cover the cuts.

The reduction stage requires a source catalog to generate a source mask for image alignment. We use \textit{Gaia} DR3 as the source catalog due to its complete coverage of the sky to magnitudes $>19\;G_{\rm mag}$. 

To assist in object identification during the detection stage, \ttt{TESSELLATE} builds a unified variable star catalog from several catalogs: the \textit{Gaia} DR3 variability catalog  \citep{gaiaDR3var} (I/358/varisum), DES RRyrae catalog \citep{Stringer2019} (J/AJ/158/16/table11), ASAS-SN variable star catalog \cite{Jayasinghe2018}, and ATLAS-VAR catalog \citep{Heinze2018}. The ATLAS-VAR catalog is accessed through MAST using the \ttt{mastcasjobs} package\footnote{\url{https://github.com/rlwastro/mastcasjobs}}. In the final joined catalog, we consider sources to be duplicates if they are within 2\arcsec\ and only retain the first instance.

\subsection{Cutout Reduction}

All cutouts are reduced using the \ttt{TESSreduce} package for \ttt{python}. This involves all the processing steps required to extract accurate photometry from \tess\ data: background modeling and subtraction, image alignment, and flux calibration, as described by \citep{Ridden2021}. A few of the more rigorous steps normally applied by \ttt{TESSreduce} are omitted to reduce compute time, such as the secondary background correlation correction. This process produces a difference-imaged datacube, where static sources are subtracted. We show a result of this reduction process in \autoref{fig:main} (left). Additionally, we save output diagnostic files, including the calculated scattered light background, position offsets, and the reference image used in subtraction.

\subsection{Transient Event Search \& Isolation}

The core of \ttt{TESSELLATE} is the detection and identification of events. The components of this source detection process are discussed in the following subsections. The low spatial resolution of \tess\ and artifacts from the scattered light background present challenges to reliably detect real sources while minimizing false detections. Detecting faint sources becomes particularly challenging as sources may only have significant flux in $\sim$3 pixels. For the detection process, we developed a two-stage process that uses information from individual differenced images in the first stage, and resultant event light curves for the second stage. 

To maximize source detection in the first stage, we have developed two independent source detection pipelines. The first pipeline, \texttt{SourceDetect}, utilizes machine learning techniques to identify sources in images (\S~\ref{sec:sourcedetect}), while the second pipeline uses conventional source detection (\S~\ref{sec:starfind}). These two methods rely on the effective PSF (ePSF) model for \tess, which is obtained through the \texttt{TESS\_PRF} package\footnote{Note: while it is common practice to refer to the \tess\ effective PSF as the PRF, we reserve that term for the intra-pixel response function.} \citep{tessprf}. Both methods are capable of detecting decreases and increases in flux which we refer to as negative and positive detections, respectively. This process allows us to identify transients, variable stars, exoplanet transits, and any other phenomena that changes the source brightness. 

In the second stage we isolate spatially co-located detections into ``objects", which are then temporally isolated into ``events". This source detection process is agnostic of light curve shape, or object type. All detections are recorded and quality cuts applied in post-processing stages. 

\subsubsection{SourceDetect} \label{sec:sourcedetect}
The size and complexity of astronomical datasets have seen huge growth in recent years due to the deployment of large-scale surveys such as ZTF \citep{Bellm2019}, PS1 \citep{Chambers2016}, ATLAS \citep{Tonry2011}, and \tess. Such changes have required astronomers to develop automated tools to analyze these datasets. Machine learning algorithms are arguably the most popular example of innovations in this area and have already been applied to areas such as detection and classification \citep[e.g.][]{Baron2019}. \texttt{SourceDetect}\footnote{\url{https://github.com/andrewmoore73/SourceDetect}}  is a novel object detection method that searches for transient and variable star events. It employs a trained convolutional neural network \citep[CNN,][]{Hinton2006,Benigo2009,LeCun2015} to identify candidate {\textquotedblleft}PSF-like{\textquotedblright} events within \tess\ images or image cutouts of any shape. 

The \texttt{SourceDetect} CNN was designed to identify both positive and negative-flux events and to differentiate between them and false objects such as overexposed regions or background artifacts. In the CNN architecture, nine parameters are extracted per detection: the likelihood that it is a real event, its dimensions, its pixel position, and its associated probability for four object classes. These classes are positive events, negative events, and two representing false objects such as background artifacts. The model was trained and tested using a dataset of 12,800, $16\times16$ pixel images with a background generated from a Gaussian distribution to mimic the real reduced \textit{TESS} images. Each image contained up to two objects from the four classes with varying brightnesses and pixel positions. Further details on \texttt{SourceDetect} will be presented in \citep{Moore2025}.

\subsubsection{Starfind} \label{sec:starfind}

In addition to \ttt{SourceDetect}, we use conventional source finding methods using the \ttt{TESS\_PRF} effective \tess\ PSF. The \texttt{Starfind} source detection method relies on the \texttt{photutils} \citep{Bradley2024} starfind algorithm. With starfind, a kernel can be specified to represent the ePSF of sources in the image. We apply this detection method to every frame in the \tess\ sector. 

As the \tess\ ePSF is under-sampled, the shape changes significantly under sub-pixel shifts. To account for the variety of appearances for the ePSF we run starfind five times with kernels that correspond to a perfectly centered source, then sources shifted by 0.25~pixels in x and y for each corner. While this combination of sources will not perfectly represent all possible \tess\ ePSF configurations, it is effective at identifying sources. We compress the results from the 5 searches into a single table, dropping co-located duplicates. 

At the detection stage we aim to identify all possible sources, so a low detection threshold of just 2$\sigma$ above the background is set. We remove obvious bad detections by cutting all detections with reported FWHM less than 0.8 and greater than 7~pixels. These FWHM cuts are effective at removing cosmic rays, and broad features, such as a region of poorly subtracted background. We also add a further cut on the standard deviation of the background as calculated by the \ttt{TESSreduce} reduction at the location of the detection. This check on the background limits contamination from poor scattered light subtractions.

This starfinding detection method outputs a table similar to that produced by SourceDetect. If both SourceDetect and starfind are run on the data, as is default in \ttt{TESSELLATE}, the two tables are joined, retaining only the first instance of detections that are spatially located and occur  in the same frame.

\subsubsection{Event Isolation} \label{sec:isolation}

Following the detection of point sources in each \tess\ FFI, we then group sources into spatially co-located ``objects",  which can then have temporally continuous ``events" associated with them. For each object, we consider negative and positive flux detections independently to create negative and positive events. As each event is separated we are able to identify and analyze independent events, such as multiple stellar flares (e.g. \autoref{fig:main} flare star example), or identify individual peaks and troughs of variable stars.  

Detections are grouped spatially using the DBscan clustering algorithm through the \texttt{scikit-learn} package \citep{scikitlearn2011}, implemented with a minimum sample number of 1 and $\rm eps=0.5$. Each object is assigned a unique ID, and is then parsed through time to isolate temporally independent events. This process builds on the method used in the Kepler/K2: Background Survey \citep{Ridden2020}, where consecutive detections are considered as a single event. To limit random noise and missing detections breaking up continuous events we consider two independent detections to be of the same event, if there are fewer than 5 non-detection frames between them. While we record all detections, we require ``events'' to be constructed from 2 or more consecutive detections to limit contamination from cosmic rays. As we treat positive and negative flux triggers independently, we obtain a series of events that cover each maxima and minima of the light curve. 

The times and durations of these events are refined through analysis of their light curves; in many cases, the undersampled ePSF of \tess\ is susceptible to being obscured by noise, while the cumulative signal in the light curve often remains significant. For each spatially resolved object, we extract light curves using the circle aperture method implemented in \ttt{photutils}, with a 1.5~pixel radius. We calculate a ``significance light curve" by calculating the signal significance of the event relative to a ``quiescence lightcurve''. We define the quiescence light curve as the light curve within 24 hours before and after the event, where points within half an hour of the event are dropped\footnote{This time duration is not changed in the case of the event occurring near a data gap.}. We then subtract the quiescent mean from the event light curve and divide by the quiescent standard deviation. With the significance light curve for each event, we can identify accurate start and end times, which we consider to be when the event signal exceeds and returns within 3$\sigma$ of the median respectively. This refines the event times such that they fully encompass the interval within which the light curve exhibits significant flux. We take the start time as the timestamp of the event.

We also use the significance light curve to define two additional detection quality metrics: the maximum and median significances of the light curve during the event. As the significance is calculated from the light curve near to the events, these metrics provide a means of removing spurious detections that are not statistically significant in the light curve, despite being identified from the images. 

At the end of this, all events are collated into a single catalog. Coordinates are calculated at the object level from an average weighted by the detection significance, and are then crossmatched against the variable star catalog described in \S~\ref{sec:var_cat}. If the object is within 10\arcsec\ of a source, they are assigned a ``Type" taken from the catalogs. This type applies to all events that correspond to an object. For each object we also use Lomb-Scargle periodograms, as implemented in \texttt{scipy} to identify prominent frequencies and their power in the light curve. 

At the end of this process we have a table containing all independent events and key parameters which can be used for event filtering. For each event we save the pipeline light curve, and generate figures to be used in \S~\ref{sec:zooniverse}.

\subsection{Targeted Search Functionality}
\label{sec:tsf}
While the primary pipeline workflow is designed as an untargeted survey, there are situations where it is beneficial to incorporate functionality for targeted searches as well. These can include attempts at cross-matching with known uncertainty regions of particular astrophysical events, such as gravitational wave detections or GRBs. As full \ttt{TESSELLATE} runs can be highly resource consuming, as outlined in \S~\ref{sec:compute}, it is convenient to quickly process smaller regions of interest.

To address this, we include an alternative \ttt{TESSELLATE} operational mode that allows for targeted observation of a known event. Taking an estimated coordinate, an associated one dimensional position uncertainty, and a time onset for the event, the pipeline determines its overlap with \tess\ imaging. If an overlap is found, the pipeline is called to process only the relevant sector and camera/CCD combinations. 

Once the run is complete, three parameters are required: the maximum time duration of the transient, the maximum time delay between the initial event time and the event onset in \tess\ (to account for delays between emission mechanisms e.g. the gamma-ray emission and optical afterglow of GRBs), and a minimum light curve significance above the local median. The light curves and information of the brightest events that meet all the criteria will then be returned.

\begin{figure*}
    \centering
    \includegraphics[width=\textwidth]{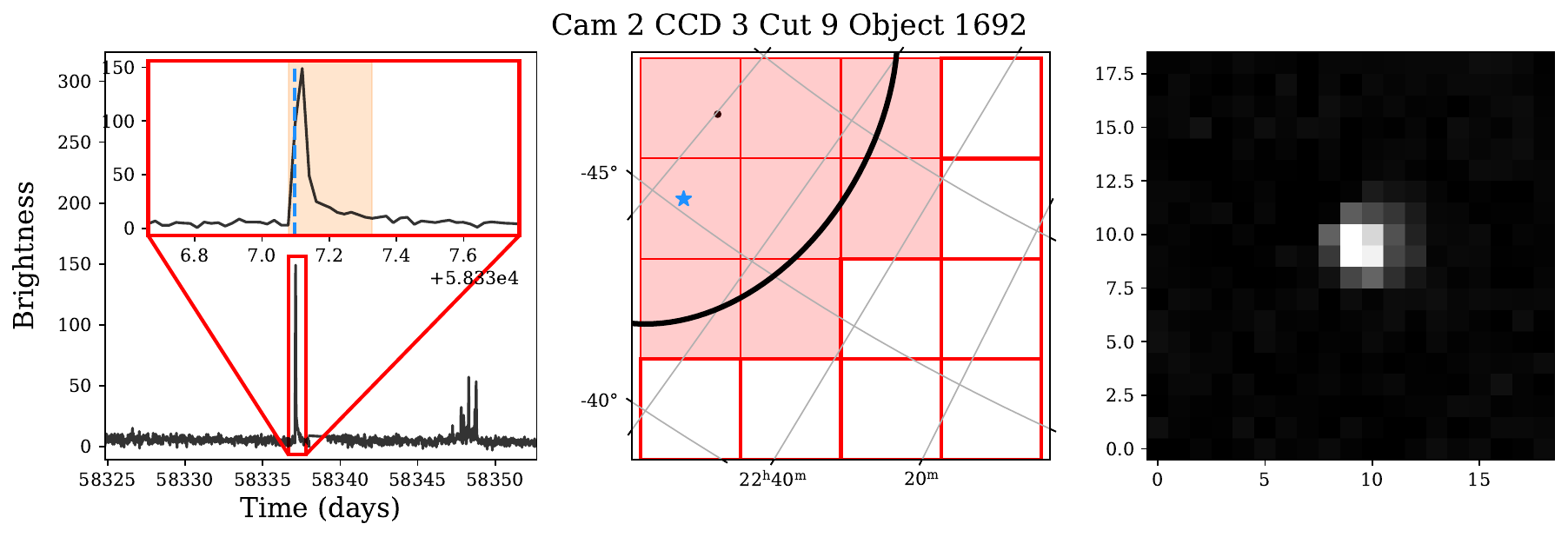}
    \caption{Targeted search for an optical counterpart to GRB180807A. \textbf{Left:} Event light curve for the detected event with the red insert zooming in on the detected transient. The transient is highlighted by the orange shaded region and the input trigger time is shown with the blue dashed line. \textbf{Center:} Plot of the spatial search area, where the central coordinates and uncertainty are shown with a black point and black line, respectively. Cuts of the \tess\ CCD that lie within the error region and are searched are highlighted in red, with the blue star representing the detected candidate. \textbf{Right:} Cutout of the difference image at the brightest point of the detected transient.}
    \label{fig:targeted}
\end{figure*}

\autoref{fig:targeted} displays an example of this mode applied to the coordinates of the known \tess\ detection of GRB180807A. This shows the most significant event that coincides temporally and occurs within the GRB's uncertainty region as reported by the \textit{Fermi} space telescope. The left panel row presents the event's light curve with respect to the GRB's reported onset time in blue. The middle panel displays the CCD upon which the candidate was found (whose location is marked by the blue star), with the red grid representing the borders of each cutout; cutouts having overlap with the black localization region of the GRB are shaded. The right panel presents the brightest frame of the event.

\subsection{Candidate Selection}\label{subsec:candidate_sel}
With \ttt{TESSELLATE} we can search all TESS data through both targeted and untargeted searches. Up to this point \ttt{TESSELLATE} uses a relaxed detection criteria to maximize detections of faint events. All detections are saved, and used to identify individual events. However, this process results in a large number of false detections which we then cut by imposing quality restrictions.

The main quality criteria include: image detection significance, light curve significance, flux sign of the event, event type, and standard deviation of the background. These options allow for high degrees of specification in the returned events, and can be chosen such that contamination is minimal. 

For the general transient search we require that the event:
\begin{itemize}
    \item is not matched to, or classified as a variable star,
    \item has an increase in flux, 
    \item has a signal to noise ratio greater than 2.5$sigma$ in the FFIs,
    \item has a mean light curve significance greater than 2.5$\sigma$ above the quiescence light curve,
    \item the standard deviation of the background of the calculated background at the position of the event is less than 50~counts.
\end{itemize}
These conditions reject most false positives while preserving faint events. The resulting sample can be cleaned of the remaining false positives through manual vetting. If stricter conditions, such as an increase in the signal significance are imposed, then contamination fraction becomes minimal at the expense of cutting faint events. Exoplanet candidates are identified through a similar procedure; however, the flux sign must be negative and the event duration must be less than a day.

\subsection{Identifying Exoplanet Candidates}
\label{sec:exoident}

Since \texttt{TESSELLATE} can detect negative events, it can detect exoplanet candidates through dips in their light curves. As transits have expected shapes, they are readily disentangled from other events. We have augmented \ttt{TESSELLATE} with a post-processing pipeline to identify exoplanet candidates. This pipeline identifies negative events, and alters the light curve by flattening and normalizing it to the reference image counts. It then inspects transit characteristics such as depth and duration. We apply quality cuts on parameters such as signal-to-noise, the number of transits detected, and source crowding.

For each sector we run the exoplanet pipeline to identify all possible exoplanet candidates which then enter the planet vetting process. We fit the transits for each candidate transit with a simple parabolic model, filtering out candidates with low $\chi^2$ values. This step removes obvious outliers, such as detections of spurious noise. Furthermore, the processed light curve must still have a significant event at the times identified by \ttt{TESSELLATE}. We then identify the primary period and phase-fold the light curve. 

We compute preliminary classifications for each phase-folded light curve with \texttt{TRICERATOPS} \citep{Giacalone2020,Giacalone2021}. 
This pipeline identifies the best-fitting transit model from a suite of possible models, covering types of eclipsing binaries and exoplanet transits. 
Alongside light curve information, \texttt{TRICERATOPS} uses stellar parameters from the \textit{TESS} Input Catalog and assess the potential impact of nearby sources. 
Candidates which are assigned high probability of being exoplanets are taken as \ttt{TESSELLATE} exoplanet candidates.

\subsection{Cosmic Cataclysms: Citizen Science with Zooniverse}
\label{sec:zooniverse}
The enormous data volume of \tess\ produces tens of thousands of candidates per sector. While most false positives are removed by our triage steps, manual vetting is required for faint candidates. To produce a clean set of transients detected by \ttt{TESSELLATE} we have created the Cosmic Cataclysms\footnote{\url{https://www.zooniverse.org/projects/cheerfuluser/cosmic-cataclysms}} citizen science project, hosted by Zooniverse. Zooniverse has hosted a wide variety of projects, and has been highly successful for transient searches such as the `Supernova Hunters' project to detect supernovae observed by the Pan-STARRS 1.8m telescope \citep{Fulton2020}.

\begin{figure}
    \centering
    \includegraphics[width=1\linewidth]{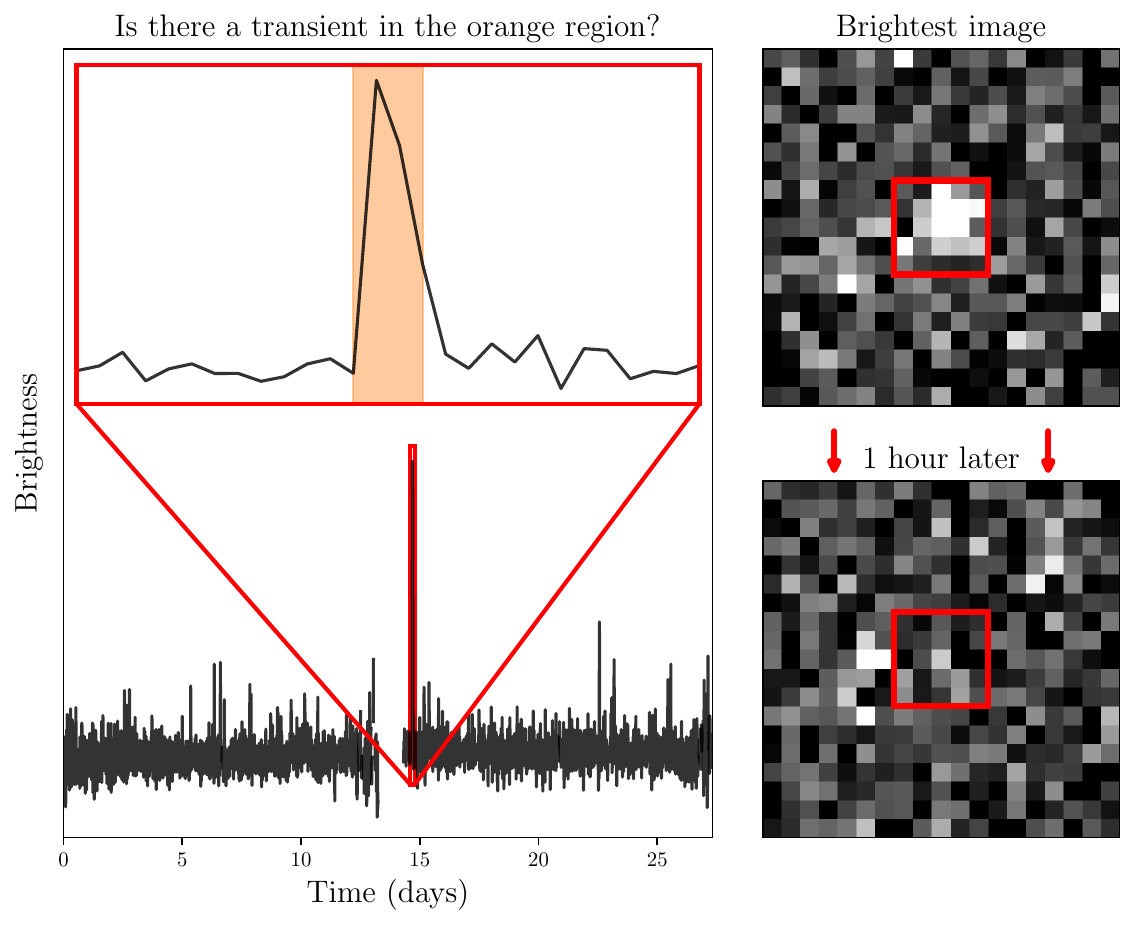}
    \caption{An example of a plot provided to the citizen scientists on Zooniverse. The upper-left shows a zoomed-in section of the difference image from the \tess\ light curve (lower-left). The upper-right displays an image from the brightest frame around the detected source, while the lower-right panel shows the frame one hour after the brightest observation. With this information citizen scientists are asked to classify the target.}
    \label{fig:zoo_example}
\end{figure}

In Cosmic Cataclysms, citizen scientists are asked to classify transient candidates based on figures such as \autoref{fig:zoo_example}. These figures contain the \tess\ light curve with the event detected by \ttt{TESSELLATE} highlighted in orange in the insert panel. We also include images showing the brightest frame centered on the event and another frame 1 hour later to identify movement of the source that might indicate that the event is an asteroid. Classifications by the citizen scientists are taken as a clean sample.

\subsection{Computational Resources} \label{sec:compute}

The \ttt{TESSELLATE} pipeline is designed to operate on SLURM based computational servers; we are primarily using the OzSTAR supercomputing node at Swinburne University. Each \tess\ sector is batched individually, with the primary processing steps spanning \S~\ref{sec:data} to \S~\ref{sec:isolation} all utilizing a single Standard Computer node with 32 cores. Each sub-process is initialized with memory and runtime limits specific to the task and sector; \autoref{tab:runtimes} displays the average requirements for 10~minute cadence data processing. These values are reduced for the 30~minute cadence data, and effectively doubled for the 200~second cadence data. As \texttt{TESSELLATE} has been constructed to be highly parallelized, multiple cuts are processed simultaneously. 

\begin{deluxetable*}{lccc}
\centering
 \tablecaption{Average processing requirements for a 10~minute cadence sector for TESSELLATE sub-processes as deployed on an OzSTAR Standard Computer node with 32 cores. The data reduction and event detection steps run independently for each cut, so here we list the memory requirement to process a single cut. We exclude the time required for downloading the \tess\ data.}
 \label{tab:runtimes}
 \tablehead{
   \colhead{Sub-process} & \colhead{Memory (GB)} & \colhead{CCD runtime (CPU hours)} & \colhead{Sector runtime (CPU hours)} }
\startdata
  Cube Data & 120 & 30 & 510 \\
  Cut Cubes & 30 & 20 & 256 \\
  Data Reduction & 110 & 770 & 12300 \\
  Event Detection & 60 & 260 & 4100 \\
 \enddata
\end{deluxetable*}

\section{Detection efficiency and event rates}
\subsection{Detection Efficiency}
\label{sec:eff-rates}

Understanding the detection efficiency of \ttt{TESSELLATE} is crucial for calculating its expected detection rate of real events. To this end, we inject simulated sources into the datacubes and compute recovery rates for each of \tess's cameras. These sources are generated by \ttt{TESS\_PRF} to match the variable ePSF of \tess, and are randomly assigned  an $(x,y)$ position on the detector (at least 3~pixels from the image boundaries), a frame number between 0 and the total number of frames, and a brightness. The brightness is drawn from a log-uniform sample, ranging from 375 to 1.5 \tess\ counts; using the average \tess\ zeropoint of 20.44 \citep{TESShandbook}, this corresponds to an $m_{T}$ of between $14$ and $20$. We apply both detection methods independently, which allows us to compare their detection efficiencies. 

\begin{figure*}
    \centering
    \includegraphics[width=\textwidth]{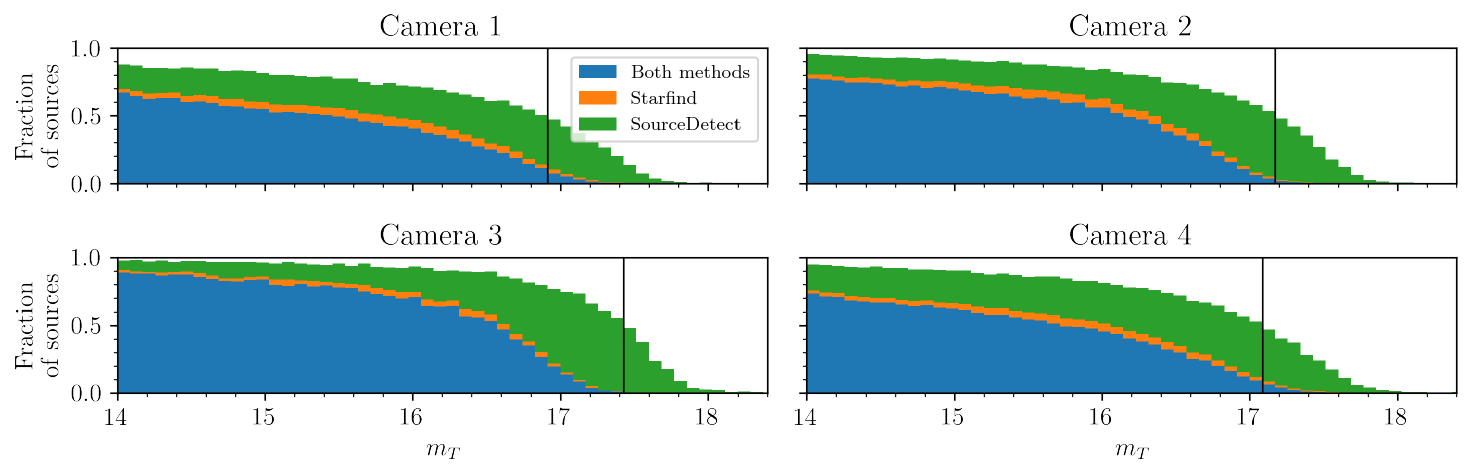}
    \caption{Mean recovery rates for the cameras in Sector 29. The recovered sources are flagged by their detection methods (Starfind and SourceDetect). The black line indicates the magnitude at which the recovery rate drops below 50\% for each camera. 
    \label{fig:recovery}}
\end{figure*}

The recovery rates for our test data for Sector~29 are shown in \autoref{fig:recovery}. From both methods we have high recovery rates for all cameras until $\sim17\; m_{T}$ at which point the recovery rate declines significantly. This decline coincides with a source brightness equivalent to a signal to noise ratio $<3$ when considering the average background. As the noise profile is not uniform across cuts, faint sources are only recoverable in areas with low noise. 

From Sector~29 we find that the cameras and CCDs are largely consistent in their recovery rates. The recovery rate depends on our ability to reliably model and subtract the scattered light background. This becomes particularly challenging for cameras with highly dynamic source scattered light backgrounds and crowding, such as cameras near the ecliptic and galactic plane. We see the impact of the scattered light background in \autoref{fig:recovery} where Camera 1 has a significantly worse recovery rate for all sources when compared to the cameras further from the ecliptic.

\begin{figure}
    \centering
    \includegraphics[width=1\linewidth]{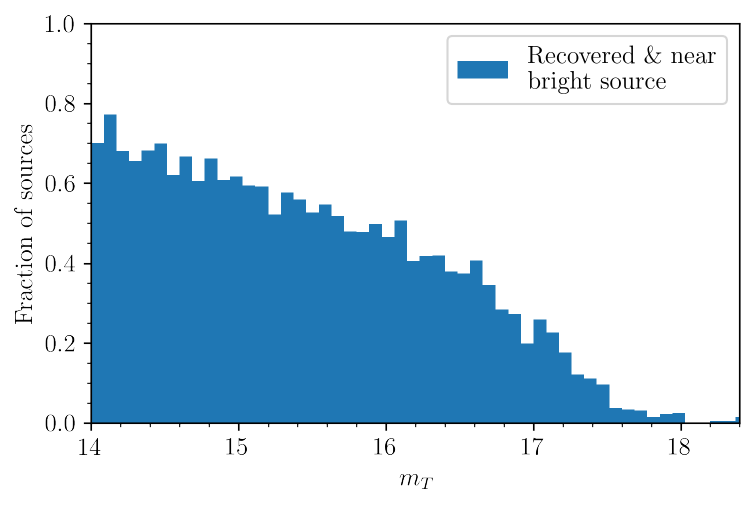}
    \caption{Mean recovery rates across Sector 29 of ePSFs injected within a one pixel radius of a bright source. These are flagged using the \ttt{TESSELLATE} source mask.   
    \label{fig:bright_source}}
\end{figure}

Source recovery is also limited by subtraction artifacts. Poorly subtracted bright sources can leave complex residuals which can obscure even bright injected events. Examples of these residuals can be seen in \autoref{fig:main} (left) which are often characterized by a checkerboard pattern of positive and negative fluxes. For Sector~29 we find that $\sim$20\% of all injected sources are considered to be close to bright sources. In \autoref{fig:bright_source} we show the recovery rate of the $\sim$20\% of sources that were injected near a bright source. While it is clear that subtraction artifacts impact the recoverability of all sources, sources fainter than $16\; m_{T}$ are significantly impacted.

The results of our efficiency trials are consistent with other work, although in some instances our detection threshold is brighter.
\citet{Fausnaugh2023} found that GRB230307A was undetectable when it reached a magnitude fainter than $m_{T} > 17.79$. 
While \ttt{TESSELLATE} is able to recover sources to equivalent depths, the low recovery efficiency suggests that there may be room to improve sensitivity in our pipeline.

\subsection{Event Rates}
\label{sec:rates}
With these detection efficiencies, we can make estimates for the expected number of individual events of known transient classes that have been observed by \tess\ and will be recovered by \ttt{TESSELLATE}. For this rate estimation we use the average recovery rate from all cameras in Sector~29 and assume it to be representative of all sectors. While \tess\ has three distinctive observing cadences, we do not see significant changes in recovery rate tests between the data. The expected survey volume for a transient is given by:
\begin{equation}
    V = \frac{\Omega}{3}\left(10^{\frac{1}{5}(m-M+5)}\right)^3,
\end{equation}
where $m$ is the apparent magnitude limit, $M$ is the absolute magnitude of the transient, and $\Omega$ is the solid angle of the Field of View. For this estimation we use the $i$, and $r$ magnitudes as approximations of the \tess\ band magnitude. 

To account for the magnitude dependent recovery rates, we calculate the number of events by summing concentric volume segments. We create the volume segments by iterating through the apparent magnitude in steps of 0.1. Each volume segment is then multiplied by the event recovery rate for its corresponding apparent magnitude range. We then sum the scaled volume segments to obtain the effective survey volume. 

Here we only consider extragalactic transients with shorter lifetimes, as this will be the primary discovery space for \tess. In particular we consider kilonovae, fast evolving luminous transients (FELTs), fast blue optical transients (FBOTs), and GRB afterglows. For the kilonovae rate, we use a value of $1540_{-1220}^{+3200} \text{ Gpc}^{-3} \text{ yr}^{-1}$, and an absolute magnitude of $-16$ based off the GW170817 light curves \citep{Abbott2017, Villar2017}. For the FELT rate, we use a value of $6400 \pm 2400 \text{ Gpc}^{-3} \text{ yr}^{-1}$, with an absolute magnitude that ranges from $-16.5$ to $-20$ \citep{Drout2014}. Using the rate derived for FBOTs similar to AT2018cow in \cite{Ho2023}, we use a volumetric rate of $0.3-420\text{ Gpc}^{-3} \text{ yr}^{-1}$ with a peak absolute magnitudes ranging from $-16$ to $-22$ mag. For simplicity during our calculations, we can set the rate to be $210.15 \pm 209.85 \text{ Gpc}^{-3} \text{ yr}^{-1}$. 

For our GRB afterglow rate calculation, we employ a more detailed approach due to the large variability that arises in the population from beaming angle effects. We adopt the population synthesis model described in \cite{Freeburn2024} which involves generating the synthetic population of GRBs described in \cite{Ghirlanda2013} based on their jet half-opening angles and Lorentz factors. \texttt{afterglowpy} \citep{Ryan2020} is then used to construct their corresponding afterglow light curves. For off-axis events, we assume a power-law jet structure with a uniform core where $\gamma$-ray emission is restricted to the core. The angular power-law index is set to 0.8 and the half-opening angle of the jet's wings are 1 radian. The peak luminosities of these light curves are then multiplied by \ttt{TESSELLATE}'s recovery rates to produce total rates for both on and off axis events. As these predictions are from a population synthesis model, the errors associated with them are simply Poisson noise.

We make a number of key simplifications in our calculations. While the observing strategy for a \tess\ sector consists of two $\sim$13~day continuous observing blocks separated by 1 day, we simplify this to 23$\;$days per sector. We also exclude CCDs from the rate calculations if they are pointing within 15$\rm ^o$ of the galactic plane. Most importantly, we do not consider the effects of \ttt{TESSELLATE}'s two-detection condition; as we are reporting rates based on the likelihood of detecting the single peak frame, many dim events are included in our rates that would be ignored by \ttt{TESSELLATE} as their second frame flux falls below the detection threshold. Without fully detailed population synthesis modeling for all transient classes, such a consideration would be highly biased by speculative expectations of the events' evolution.

\begin{deluxetable*}{lccccc}
 \tablecaption{Estimated event rates for kilonovae, FELTs, FBOTs, and GRB afterglows for Sector~29, and extrapolated to include all sectors from S01 to S78. For FELTs and FBOTs we list the expected range of absolute magnitudes. While \tess\ has a relatively bright limiting magnitude, the total survey volume make it likely for \ttt{TESSELLATE} to recover many fast extragalactic transients observed by \tess.
 \label{tab:volrate}}
 \tablehead{
\colhead{Sector} & \colhead{Kilonova (NS-NS)} & \colhead{FELT} & \colhead{FBOT} & \multicolumn{2}{c}{GRB}\\
     \colhead{}    & \colhead{} & \colhead{$\left(-16.5\,\rightarrow\, -20\right)$} & \colhead{$\left(-16\,\rightarrow\, -22\right)$} & \colhead{On-axis} & \colhead{Off-axis}}
  \startdata
  S29 & $0.03^{+0.06}_{-0.02}$ & $0.23 \pm 0.09$ $\rightarrow$ $29 \pm 11$ & $0.004 \pm 0.004$ $\rightarrow$ $15 \pm 15$ & $5 \pm 3$ & $14\pm4$ \\
  S01-S78 & $2^{+4}_{-2} $ & $17 \pm 6$ $\rightarrow$ $2102 \pm 788$ & $0.3 \pm 0.3$ $\rightarrow$ $ 1093 \pm 1092$ & $381\pm20$ & $1026\pm32$
 \enddata
\end{deluxetable*}

The recovery rates are shown in \autoref{tab:volrate}, where we estimate the number of events for Sector~29, and extrapolate to all observed sectors, using the Sector~29 recovery rates. In this simple estimation, our Kilonova rates are consistent with \citet{Mo2023}, however, it is likely that we overestimate the number of events that \ttt{TESSELLATE} will recover. Regardless, this does indicate that \tess\ is likely to have observed a significant number of known faster transients which currently lie undiscovered in archival \tess\ data, even if just a single frame is present. As we do save single frame detections, studies on these cases will be conducted once a large population is obtained. Furthermore, after searching each sector of \tess\ data, we will be able to conduct detailed analysis of event recovery to create strong constraints on the rates of fast extragalactic transients in the local universe.

\section{Science cases}
\label{sec:cases}

Since \ttt{TESSELLATE} is an untargeted survey of the variable sky, we will detect phenomena that span many areas of astronomy. 
Our primary focus is fast extra-galactic transients.
However, there are several science cases for which \tess\ data and \ttt{TESSELLATE} detections can provide significant benefit: we present a range of examples here.

\subsection{Fast Transients}

The primary focus of the \ttt{TESSELLATE} survey is to discover new transients. 
This designation covers all events which are one-off or irregular in nature, including GRB afterglows, classical and dwarf novae, gravitational microlensing events, kilonovae, and other not-yet-cataloged fast phenomena. 
Longer-duration transients, such as supernovae, are outside the scope of this survey; while detectable by \ttt{TESSELLATE}, these events are routinely discovered from the ground and exist outside of the unique discovery space for \tess. 
Therefore, we do not discuss them here.

\subsubsection{GRB Afterglows}

To date, multiple GRB afterglows observed by \tess\ have been documented: \citet{Smith2021} presents the first GRB lightcurve observed by \textit{TESS}; \citet{Roxburgh2023} finds 8 afterglow candidates from 79 poorly-localized GRBs; \citep{Jayaraman2024} analyses the \textit{TESS} light curves of 8 well-localized GRBs and identifies prompt emission in 3; \citet{Perley2025} identifies a GRB orphan afterglow observed by \textit{TESS}. These high-cadence observations offer valuable insights into the physical processes that power GRBs. For example the mechanism which produces the prompt emission reported in \citet{Jayaraman2024} is not well understood. Furthermore, \citet{Roxburgh2023} identifies a delay time between GRB trigger and afterglow onset which is attributed to the travel time required for the GRB to excite the circumstellar material. Further high-cadence observations are needed to build a clearer understanding of GRBs and their afterglows, a role which \tess\ is poised to fill. Simple rate calculations, such as those in \S~\ref{sec:rates} suggest that there may be many more in archival data.

A key science case for \ttt{TESSELLATE} is the search for ``orphan afterglows".  While a GRB is a highly collimated jet with a small opening angle, optical afterglows are emitted at a wider angle and thus can be viewed off-axis. Since \tess\ has the time resolution to detect afterglows, and covers a large volume, it is possible for our untargeted search to detect these events.

Contamination from M-dwarf flares pose a challenge to GRB afterglow surveys. These two phenomena can follow very similar light curve profiles, as shown by the similarity between the GRB afterglow and the first flare from the flare star in \autoref{fig:main}. It is particularly difficult to disentangle the two populations in monochromatic data such as \tess; thus, comparison with Legacy photometry is essential for the verification and characterization of afterglow candidates. 

\subsubsection{Cataclysmic Variables}

Cataclysmic variables (CVs) are systems containing a primary white dwarf star accreting matter from a low-mass binary partner. These systems are known to outburst in a variety of ways, arising from different mechanisms related to the depositing of matter via an accretion disk. Most commonly, these outbursts manifest as episodic increases in luminosity that last for up to a week, known as classical or dwarf novae 
\citep[for a review, see][]{Warner1986}. 

High-cadence observations, such as those provided by \tess\ are essential for understanding the nature of CVs. In the case of dwarf novae, rapid oscillations can be present in super-outbursts that are linked to the orbital period and the binary mass ratio. Previous studies of dwarf novae with high cadence data from \textit{Kepler} and \tess\ have resulted in discovery \citep{Barclay2012,Ridden2019,Boyle2024}, and the characterization of known dwarf novae \citep{Kato2013,Kato2013a,Bruch2022,Wei2023}.

A significant discovery space still exists for CVs, particularly for high cadence photometry. Even the sample of nearby CVs that are within 150~pc is predicted to be only $77\pm10$\% complete \citep{Pala2020}. Due to the large survey volume and cadence, \tess\ is an ideal instrument for discovering new out-bursting CVs, such as the previously unidentified CV in \autoref{fig:main}. With \ttt{TESSELLATE} we will add to the out-bursting CV population with high temporal resolution.

\subsubsection{Microlensing Events}

Gravitational microlensing events have been widely observed by dedicated surveys such as the Optical Gravitational Lensing Experiment \citep[OGLE,][]{Udalski1992,Udalski2015}. Rare chance alignments between background sources and foreground dark or non-emitting celestial bodies, such as free-floating planets (FFPs) or compact objects, can significantly magnify the light emitted by the background source. 
FFPs lensing events present two challenges for observation in that the magnification is small and will occur over a timescale of hours to days. High cadence and wide area surveys, such as \ttt{TESSELLATE} have the capability of detecting even the fastest lensed events.

To date, only one FFP lensing event has been proposed from \tess\ data \citep{Kunimoto2024}. This interpretation of the event, which occurred in  in Sector 61, has been challenged by \citet{Mroz23} who suggest it is likely a symmetric stellar flare. \citet{Yang2024} performed a detailed occurrence rate calculation which suggests $\mathcal{O}(1)$ FFPs will exist in \tess\ data by the end of its second extension mission in 2025. This calculation was based on background sources brighter than 16th magnitude; the untargeted nature of our survey allow for dimmer stars as background sources, though with \tess's low spatial resolution and with the maximum event likelihood region towards the galactic bulge, source crowding will present a significant challenge.

Larger lens masses, eg. brown dwarfs or compact objects, result in much longer microlensing durations. As each \tess\ sector lasts around 27 days, our survey will be sensitive to events that are weeks long. However, for timescales greater than 27 days, it will be difficult to distinguish the peak of a microlensing event from that of a long period variable star. 

\subsubsection{Kilonovae}

Kilonovae are among the most highly sought after transient phenomena, with fewer than a dozen confirmed and candidate events documented in the literature - none of which have observed rises \citep[for recent reviews, see][]{Metzger2020,Troja2023}. They arise during merger events between compact objects, eg. NS-NS or NS-BH; as these are also sources of detectable gravitational wave emission, kilonovae are highly desirable for multi-messenger event studies.

While current expectations are that kilonovae signatures are highly variable in their colors and luminosities, they are inherently dim events. This poses an issue for \tess's capability to detect them; transients peaking fainter than 18th magnitude are largely indistinguishable from noise. Thus, we find it highly unlikely that this survey will detect any kilonovae except under extraordinary conditions. 

\subsection{Variable stars}

\texttt{TESSELLATE} is also a powerful tool for discovering and documenting variable stars. Through the high-cadence observations of \tess\ we can distinguish key classes of variables using their light curves alone, even for fast variables such as $\delta$~Scuti stars. While we cross reference detections with multiple variable star catalogs, as outlined in \S~\ref{sec:sourcedetect}, we find numerous uncataloged variable stars in our early results.

\begin{figure*}
    \centering
    \includegraphics[width=1\textwidth]{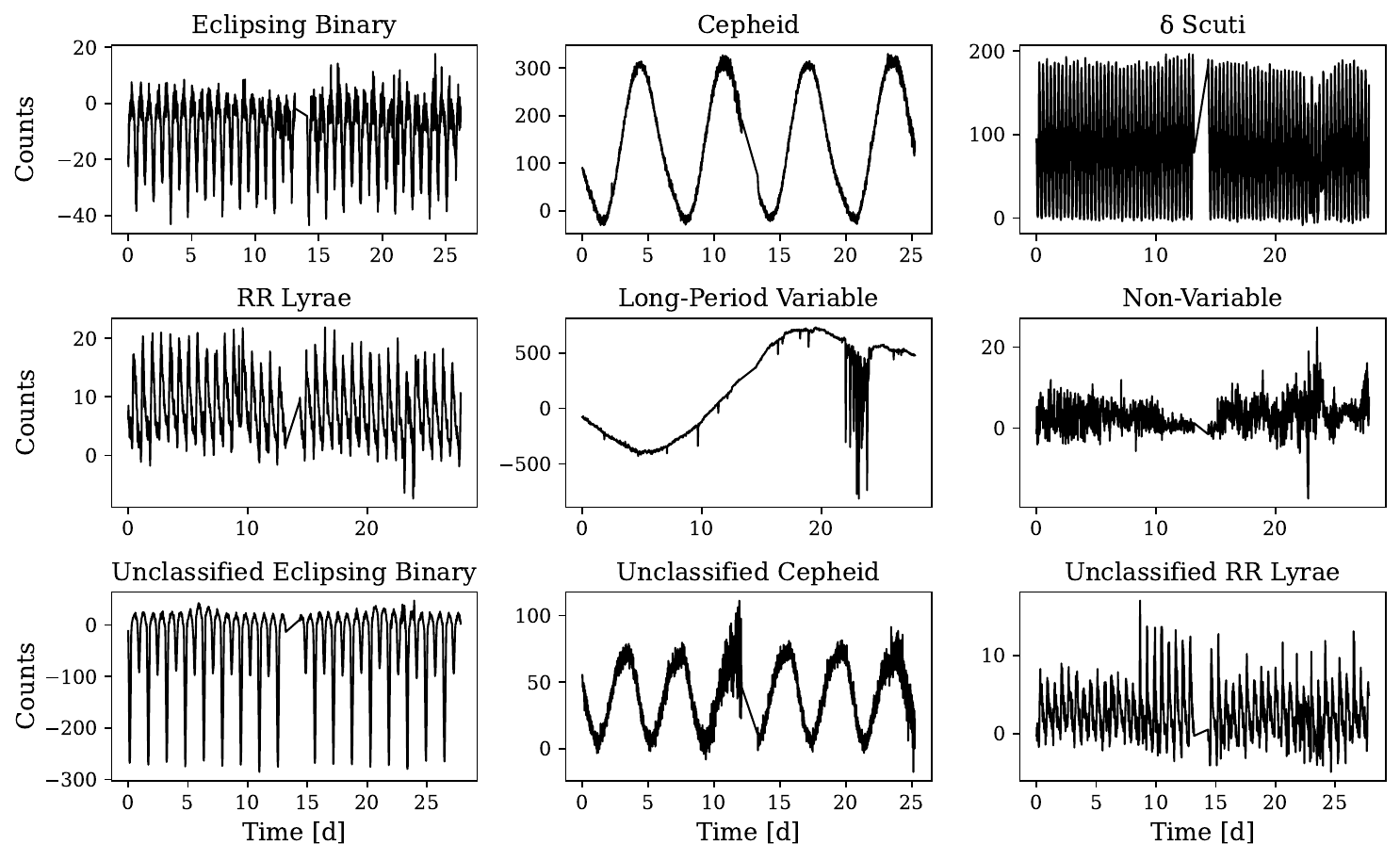}
    \caption{The light curves of eight variable stars characterized by the \texttt{SourceDetect} RFC. The top and middle rows contain previously discovered variable stars with classes matching the classifications made by the \texttt{SourceDetect} RFC. The bottom row has previously unclassified objects with variable classifications made by \texttt{SourceDetect}.}
    \label{fig:VariableExamples}
\end{figure*}

Following recent work, we can assign classifications to these stars through machine learning techniques \citep{Audenaert2021,Elizabethson2023}. \ttt{TESSELLATE} performs such a classification during the detection stage  using \ttt{SourceDetect}'s random forest classifier (RFC), which is trained on \tess\ light curves of cataloged variable stars. \autoref{fig:VariableExamples} displays several examples of known variable stars, alongside three new classifications made with the RFC. The details of the variable star classifier will be discussed in \citet{Moore2025}.

\subsection{Minor Planets}
\label{sec:asteroids}

\tess\ has an established history of bright $m_r < 20$ minor planet detection and population analysis \citep{Pal2018,Pal2020,Woods2021, McNeill2023}.
The high-cadence light curves enable shape fitting for asteroids \citep[e.g.][]{Humes2024} and characterization of comets \citep[e.g.][]{Ridden2021b}.
\ttt{TESSELLATE} provides asteroid detections as part of its transient pipeline; an example is shown in \autoref{fig:main} (lower center).

As \ttt{TESSELLATE} has a shallow magnitude limit of $\sim$17, only known minor planets will be recoverable.  We identify all minor planets in a cut through \ttt{SkyBoT} queries \citep{Berthier2006} and build ephemerides for all targets brighter than $ m_T=18$ using information from the Minor Planet Center. Asteroid positions are cross referenced against \texttt{TESSELLATE} detections, which are then removed from the transient sample. 

\begin{figure}
    \centering
    \includegraphics[width=\linewidth]{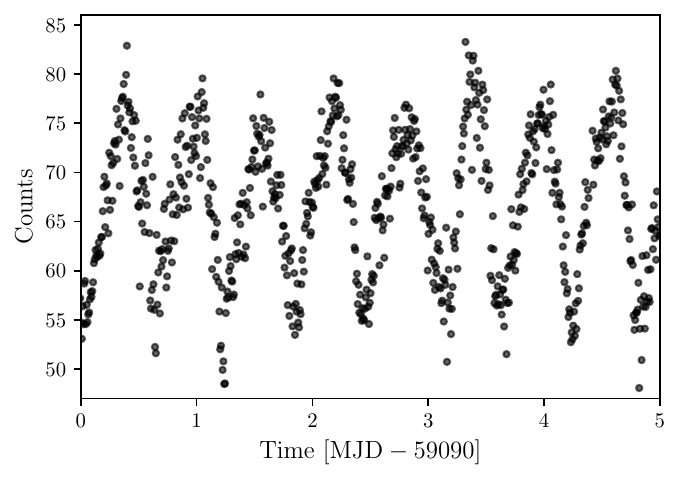}
    \caption{A section of the 10~minute cadence light curve of asteroid (5254) Ulysses from sector 29.}
    \label{fig:simple_asteroid}
\end{figure}

Although \ttt{TESSELLATE} only provides discrete detections of minor planets, we can perform forced photometry to extract lightcurves of all known objects. We stitch together lightcurves of asteroids that cross multiple cuts. The resulting high-cadence lightcurves of minor planets can be used to analyze object morphology and rotation/tumble periods. An example lightcurve segment for the asteroid (5254) Ulysses from Sector 29 is shown in \autoref{fig:simple_asteroid}. With the Sector 29 light curve we identify a rotation period of $28.71\pm0.01\;$hours, which is consistent with the asteroid Lightcurve Database \citep[LCDB 2023 release,][]{Warner2009}. Population and individual target analysis with \ttt{TESSELLATE} will be released in upcoming work.

\subsection{Flare Stars}

Stellar flares are among the most common ``transient" phenomena, and are the most numerous events in our sample. While they are contaminants for our fast extragalactic transient search, flare stars are valuable for understanding stellar processes, and atmospheric retention on exoplanets \citep{Kowalski2024}.

The complex nature of stellar flares produces a variety of different light curve profiles. While simple isolated flares appear as simple exponential declines, multiple flares can stack producing complex structures. For example, the flare star presented in \autoref{fig:main} exhibits first a simple and then a complex flare. 

As flares evolve over minute timescales, high-cadence observations are required to study flare properties. While flares can be described with simple analytical models, only the 200~second cadence FFIs are fast enough to accurately recover flare parameters \citep{Howard2021,Mendoza2022}.

With \ttt{TESSELLATE} we can construct an all-sky catalog for flare stars. Once we process the 200~second data, we can also perform population statistics on flare properties.

\subsection{Transiting Exoplanets}

\begin{figure*}
    \centering
    \includegraphics[width=\textwidth]{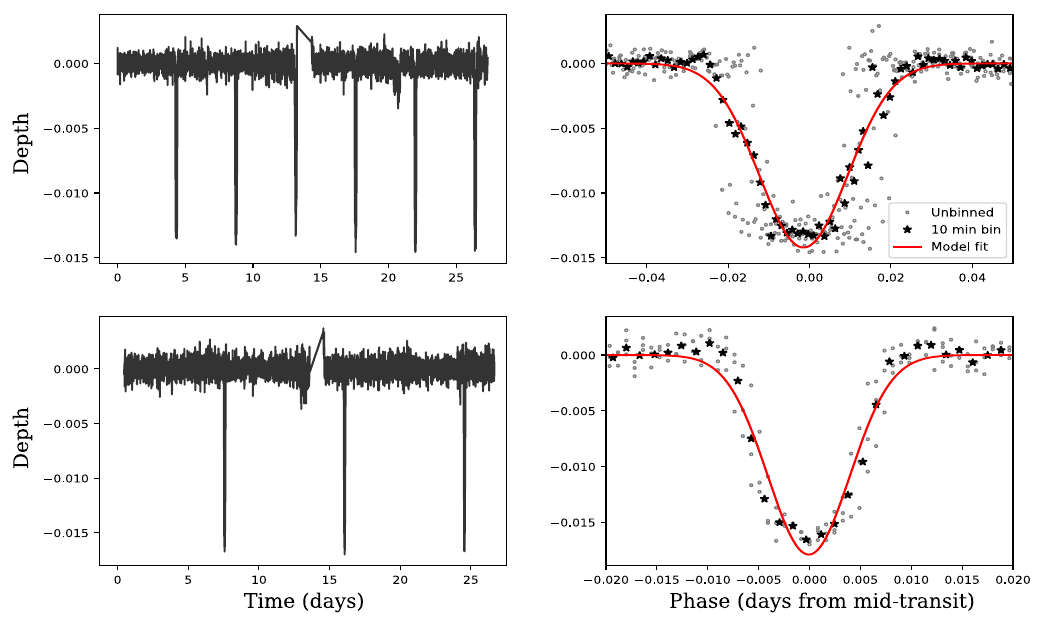}
    \caption{The light curves of known exoplanet host star WASP-62 (\textbf{Top Panels}) and a previously undiscovered exoplanet candidate TSS J0547-6921 (\textbf{Bottom Panels}). The left panels are replotted in their respective right panels after being phase-folded and fit with a standard Gaussian model (red line) which is used in the \ttt{TESSELLATE} exoplanet vetting criteria.}
    \label{fig:ExoplanetExamples}
\end{figure*}

Through our untargeted search, we are also able to detect transiting exoplanets. The post-processing pipeline that we discuss in \S~\ref{sec:exoident} has recovered known transiting exoplanets and produced tens of new candidates just from Sector 29 data. While other pipelines have conducted exoplanet searches of FFI data, they often impose magnitude limits. For example, the search described in \citet{Barclay2018} only considers stars with $I$ magnitudes $\leq 11$ for sun-like stars and $\leq 15$ for M-dwarfs. With \ttt{TESSELLATE} we impost no magnitude limit, and therefore are able to explore a new parameter space for identifying exoplanet candidates of faint stars.

For Cameras 3 and 4 in Sector 29, we recover 8 out of 27 known exoplanet host stars. Due to our search method, all known exoplanet candidates that we recover must have significant dips, therefore, we are unlikely to recover candidates with high contrast between stellar and planetary radii. In \autoref{fig:ExoplanetExamples} we show example light curves of the recovered exoplanet WASP-62 and a \ttt{TESSELLATE} exoplanet candidate TSS J0547-6921.

As many of the host stars for \ttt{TESSELLATE} exoplanet candidates are faint, follow-up and confirmation will be challenging. As with transient detections, we will make all exoplanet candidates and eclipsing binaries public as part of the \tess\ community target of interest program.

\section{The TESSELLATE Sky Survey}
\label{sec:skysurvey}

\tess\ has observed $>90$\% of the sky during its six years of operation, with most areas receiving multiple visits. With \ttt{TESSELLATE}, we can build an extensive catalog of variable sources that cover a broad range of astronomy. Following the vetting process outlined in \S~\ref{subsec:candidate_sel}, we collate all targets and build the \ttt{TESSELLATE} Sky Survey (TSS).

The catalog will report all detected transient objects, including: flare stars, variable stars, exoplanet candidates, and all other transient phenomena. Minor planet detections will not be included in the catalog. These components each pass through their own vetting and classification processes:  variable stars through the \texttt{SourceDetect} RFC, transient candidates through Zooniverse vetting, and negative-flux detections through the exoplanet pipeline.

The TSS catalog will be publicly available to support open science. Sources within the catalog are named with a convention that is dependent on their spatial and temporal coordinates. The base format we adopt for events is TSS~JHHmm$\pm$DDddTttttt where TSS is the survey designation, HHmm is the hours \& minutes components of the R.A., DDdd is the degrees and arcminutes component of the declination, and ttttt is the rounded MJD of the event. Sources that have numerous events per sector, such as active flare stars and variable stars, will follow a modified naming structure where the time component is dropped. Examples of this naming convention can be seen for the example light curves in \autoref{fig:main}.

Once events are named,  we build catalog profiles by cross referencing with existing source and transient catalogs. As the positional error for \ttt{TESSELLATE} events is $\sim$10$\rm ''$, multiple sources can be present inside the error region, complicating host identification. In such cases, we identify the most likely host though considering host position and color, which is then weighted by the morphology of the event light curve. 

TSS data is processed per sector; we are starting with the 10~minute cadence data from Extended Mission 1, spanning sectors 27 to 55. This data finds a middle-ground between temporal resolution and data processing time. We will present the first catalog once data from the southern hemispheres has been processed.

\section{Conclusion}
\label{sec:conclusion}
We present \ttt{TESSELLATE}, the first pipeline capable of differencing and extracting transient events from all data stored in \tess\ FFIs. Using the \ttt{TESSreduce} difference imaging pipeline and a range of source detection techniques, the pipeline identifies variable sources in every frame across the entire detector. Since \ttt{TESSELLATE} is an untargeted search, there are no requirements placed on event or source type. It thus has the ability to discover new transient classes, new variable stars, and even new exoplanet candidates. These sources are compiled into a table of event candidates, which is subject to various quality cuts to produce a sample of variable and transient events.

Our pipeline is currently capable of detecting and distinguishing between variable stars, exoplanet candidates, and transients. These three main event classes are processed with specific vetting and characterization pipelines. Variable stars are classified by applying a random forest classifier to the light curves with the \texttt{SourceDetect} package. Exoplanet candidates are classified where possible with \ttt{TRICERATOPS}, however, many of our exoplanet candidates have faint host stars with insufficient data to reach a classification. Due to the complex and uncertain nature of fast transient lightcurves all \ttt{TESSELLATE} candidates are vetted by citizen scientists with the Cosmic Cataclysms project hosted by Zooniverse. Already we have thousands of transients classified through Cosmic Cataclysms, with which we are building a large labeled set of \tess light curves which will be used to train a generalized fast transient classifier.

Events that pass our vetting criteria are included in the \ttt{TESSELLATE} sky survey catalog, which will contain information on likely hosts identified from deep imagining source catalogs with high spatial resolution. The catalog and pipeline light curves will be made open access for each of the two-year \tess\ missions; the first data release will be produced using the 10~minute cadence data from Extended Mission 1 covering sectors 27--55. Alongside supporting a multitude of science cases, this catalog could be cross-referenced by other surveys to readily remove repeating galactic transients such as flare stars. 

The large search volume covered by \tess\ at high cadence allows us to search for unique rapid transients. Of the known fast transients we find it is unlikely that a kilonova will be observed by \tess, however, we expect a high number of FELTs, FBOTs and GRB afterglows to be present in the data. Since we do not impose any restrictions on the light curves of transient events, we will also search for new classes of fast transients. Alongside building a library of high cadence light curves for these events, we will be able to place strong constraints on their occurrence rates. 

\tess\ observations now tile the sky multiple times, enabling an exploration of the rapid time domain. This coverage will only improve with \tess's Extended Mission 3 (EM3), which is expected to begin in September 2025. EM3 will increase the temporal baseline with longer sectors with significant overlap at mid-latitudes. \ttt{TESSELLATE} is perfectly poised to take advantage of this extended monitoring, piecing together the variable sky to better understand our dynamic Universe.



\begin{acknowledgments}

H.R. is supported by an Australian Government Research Training Program (RTP) Scholarship. R.R.H., A.M, C.M, and Z.G.L. are supported by the Royal Society of New Zealand, Te Ap\={a}rangi through the Marsden Fund Fast Start Grant M1255. R.R.H is also supported by the Rutherford Foundation Postdoctoral Fellowship RFT-UOC2203-PD. 

M.T.B. appreciates support by the Rutherford Discovery Fellowships from New Zealand Government funding, administered by the Royal Society Te Ap\={a}rangi.

The material is based upon work supported by NASA under award number 80GSFC24M0006.

We thank the Zooniverse beta testers for Cosmic Cataclysms, in particular the users: Marcossilva, Wirg78, nathancbell, Kutchkutchela, InoSenpai, skraczek, As1310, Mayahn, jklingele, Gammapat, julienlg, remaap24, cesanjerry, captaincat, Barbalbero, andre.andrade, Enriquepa, jrukmini, cosmicTaryn, mprasad, Brbljusa, Tamerciftci; and project translators: InoSenpai, aglr, Michael\_Moons-654, Cledison Marcos da Silva.

This work was performed on the OzSTAR national facility at Swinburne University of Technology.
The OzSTAR program receives funding in part from the Astronomy National Collaborative Research Infrastructure Strategy (NCRIS) allocation provided by the Australian Government, and from the Victorian Higher Education State Investment Fund (VHESIF) provided by the Victorian Government.

\end{acknowledgments}

\vspace{5mm}
\facilities{\tess, OzSTAR}

\software{\ttt{TESSELLATE},
\ttt{SourceDetect} \citep{Moore2025},
\ttt{TESSreduce}
\citep{Ridden2021},
\ttt{afterglowpy} \citep{Ryan2020},
\ttt{astropy} \citep{Astropy2013,Astropy2018,Astropy2022},
\ttt{astrocut} \citep{astrocut},
\ttt{astroquery} \citep{astroquery},
\ttt{scipy} \citep{2020SciPy-NMeth},
\ttt{photutils} \citep{Bradley2024},
\ttt{pandas} \citep{mckinney-proc-scipy-2010,reback2020pandas},
\ttt{numpy} \citep{numpy},
\ttt{matplotlib} \citep{Hunter2007},
\ttt{NIFTY-LS} \citep{Garrison2024},
\ttt{scikit-learn} \citep{scikitlearn2011},
\ttt{TRICERATOPS} \citep{Giacalone2020, Giacalone2021},
\ttt{TESS\_PRF} \citep{tessprf}
}

\bibliography{bibliography}{}
\bibliographystyle{aasjournal}

\end{document}